\documentclass[3p,times,twocolumn]{elsarticle}
\usepackage{subfig}    
\usepackage{float}     

\usepackage{ecrc}

\volume{00}

\firstpage{1}

\journalname{Nuclear and Particle Physics Proceedings}

\runauth{}

\jid{nppp}

\jnltitlelogo{Nuclear and Particle Physics Proceedings}

\usepackage{amssymb}

\usepackage[figuresright]{rotating}

\begin{document}

\begin{frontmatter}
	
	
	
	\dochead{}
	
	\title{A feasibility study of the reflection readout method of Resistive-Plate Chambers}
	
	
	
	\author[a,b]{Y.X. Ding}
	\author[a,b]{X.Y. Xie}
	\author[a,b]{J.X. Li}
	\author[a,b]{K.L. Han}
	\author[a,b]{Y.J. Sun\corref{cor1}}
	\ead{sunday@ustc.edu.cn}
	\cortext[cor1]{Corresponding author}
	\address[a]{State Key Laboratory of Particle Detection and Electronics, University of Science and Technology of China, Hefei 230026, China}
	\address[b]{Department of Modern Physics, University of Science and Technology of China, Hefei 230026, China}

	\begin{abstract}
		The conventional readout method of the RPC detector uses two sets of orthogonal readout strips placed at the both sides of the gas gap to collect signals of opposite polarities to obtain space points. A new readout method utilizing the reflected signals is proposed which only requires one set of readout strips. The reflection readout method utilizes the differences in the arrival time of the direct and reflected signals to determine the hit position. Customized transmission cables are introduced to extend the propagation distance of reflected signals to ensure sufficient separation of the two signals. Because only one side of the readout panel is connected to the FEE boards, reflection readout method can increase the geometrical acceptance and save the readout channels. Experimental setup and test results of this novel readout method is shown in this paper. It is demonstrated that a spatial resolution of sub centimeter can be achieved by processing the rising edges of the original and reflected pulses  with commonly used  electronics. 
	\end{abstract}
	
	
\end{frontmatter}


\section{Introduction}
\label{}
Resistive Plate Chamber (RPC)~\cite{R_1981}  is widely used for the detection and measurements of muons in high energy physics (HEP) experiments thanks to their high efficiencies, fast responses and relatively low cost to construct a large-area detector. The traditional readout method of RPC uses two sets of orthogonal readout strips to collect signals of opposite polarities. Each set of readout strips read outs the 1-dimensional profile of the avalanches. Since front end electronics (FEE) boards are placed at the edges of both readout panels, the geometrical acceptance of the RPC assembled by this method is limited. Therefore, in recent years, another readout method called double-end method is introduced~\cite{Li_2021}. Signals of the same polarity are read with the FEE boards equipped at both ends of one set of readout strips. The incident position along the strip is reconstructed by the difference in the arrival times of the signals at the two ends. The double-end readout method increases the geometrical acceptance and reduces the number of electronic channels. Inspired by the double-end readout method, a novel readout scheme, referred to as reflection readout method, is proposed exploiting the signal reflection. The reflected signal from the one open end of the strip propagates towards the other end that is equipped with FEE boards. The direct signal and the reflected signal will be both read out with the same electronics board in sequence. The interval between the ToAs of the direct and reflected signal can be used to infer the hit position along the readout strips. This method using only one group of FEEs increases the geometrical acceptance further. In this paper, a demonstration of this novel method and test results are reported.

\begin{figure*}[!h]
	\centering
	\subfloat[]{\includegraphics[width=2.3 in]{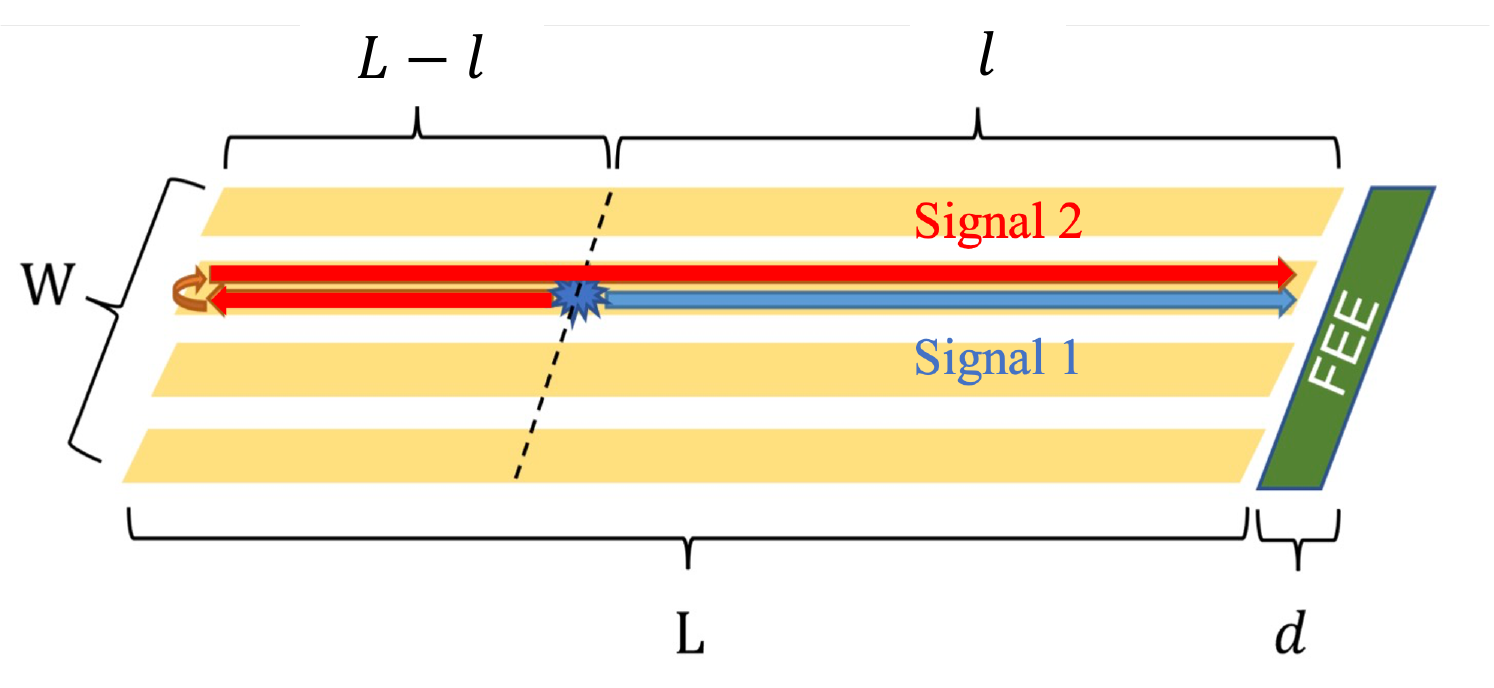} \label{X2}}
	\qquad
	\subfloat[]{\includegraphics[width=2.7 in]{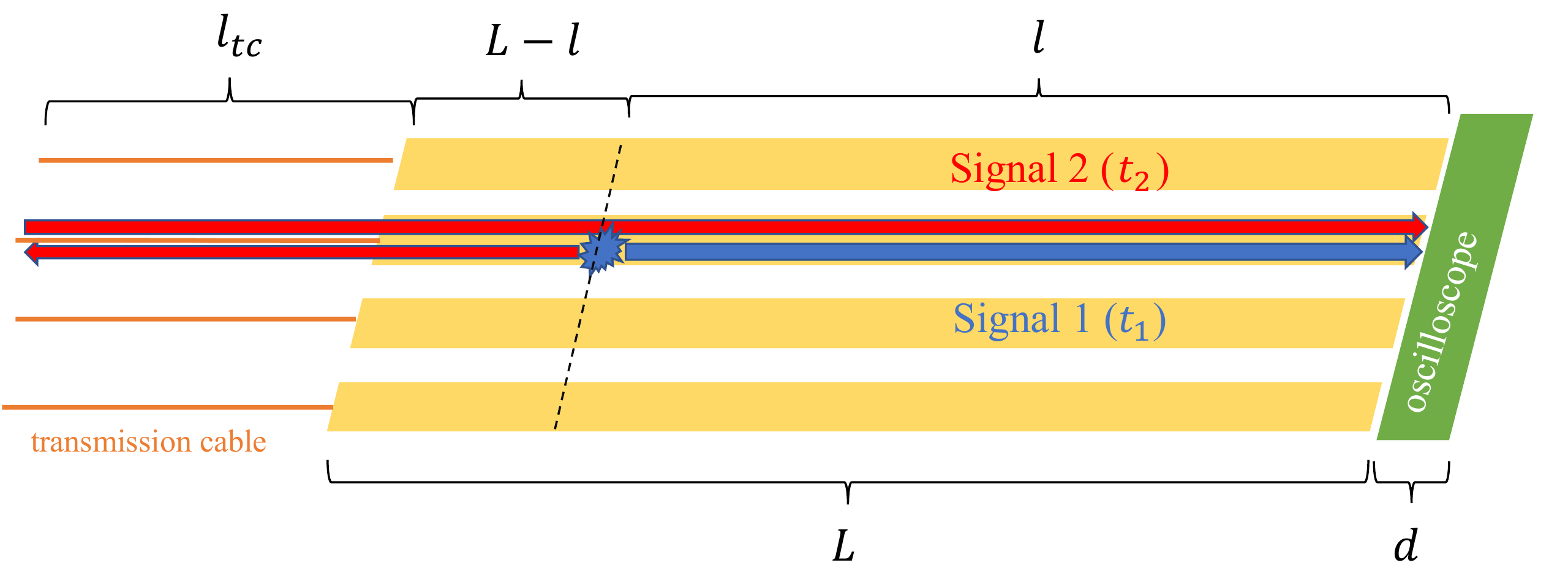} \label{Y2}}
	\caption{(a) Two induced signals propagates towards two ends of the readout strips. Signal 2 (reflected signal, in red) is fully reflected at the left end without matching resistor, and recorded by the same FEE board which has recorded Signal 1 (direct signal, in blue). (b) Optimized scheme of the reflection readout method. 'Dead area' can be eliminated by connecting transmission cables to the readout panel.}
	\label{fig:label1}
\end{figure*}

\section{Scheme optimization}
\label{} 
As shown in Figure \ref{fig:label1}\subref{X2}, when an incident particle passes through the RPC, two signals are induced on the readout panel and propagate towards the opposite direction. The direct signal (referred to as Signal 1) propagates toward the right end of the readout panel and is recorded by the FEE board. The reflected signal (referred to as Signal 2) is fully reflected at the left end without matching resistor installed, and propagates again towards the right end. Both signals are recorded by the same FEE board.

For the scheme in Figure \ref{fig:label1}\subref{X2}, there is an obvious limitation – the existence of ‘dead area’ near the reflection end where the two signals should be overlapped and can not be distinguished by the FEE. The length of ‘dead area’ can be calculated by  
$$l_{dead}=v\times F\, /\, 2,$$ where $v$ and F are respectively propagation velocity and FWHM of the RPC signals. According to a typical RPC signal~\cite{Xie_2021} with $210\,mm/ns$ of propagation velocity and $1.5 \, ns$ of FWHM, the length of 
‘dead area’ is about 157.5mm.

A possible solution is to introduce the transmission cables acting as the delay line to increase the interval between the ToAs of the original and reflected signal (referred to as time difference). The optimized design is shown in Figure \ref{fig:label1}\subref{Y2}. One end of the transmission cable is connected to the left end of readout panel while the other end is electrically floating to ensure full signal reflection. These transmission cables are customized with special characteristic impedance of 20 $\Omega$ which is the same as the impedance of readout strips. In this way, Signal 1 can be smoothly transmitted between the transmission cables and the readout panel, and the dead area is highly suppressed at the same time. The hit position along the readout strips $l$ can be reconstructed by $$t_{diff} = 2(L-l)/v + t_{lc},$$where $t_{diff}$ and $v$ are the time difference of two signals and the propagation velocity of the signals along the readout strips, and $L$ and $t_{lc}$ are the length of the readout strips and the transmission time of the signals along the transmission cables.

\section{Experimental setup}

\begin{figure*}[!h]
	\centering
	\subfloat[]{\includegraphics[width=2.3 in]{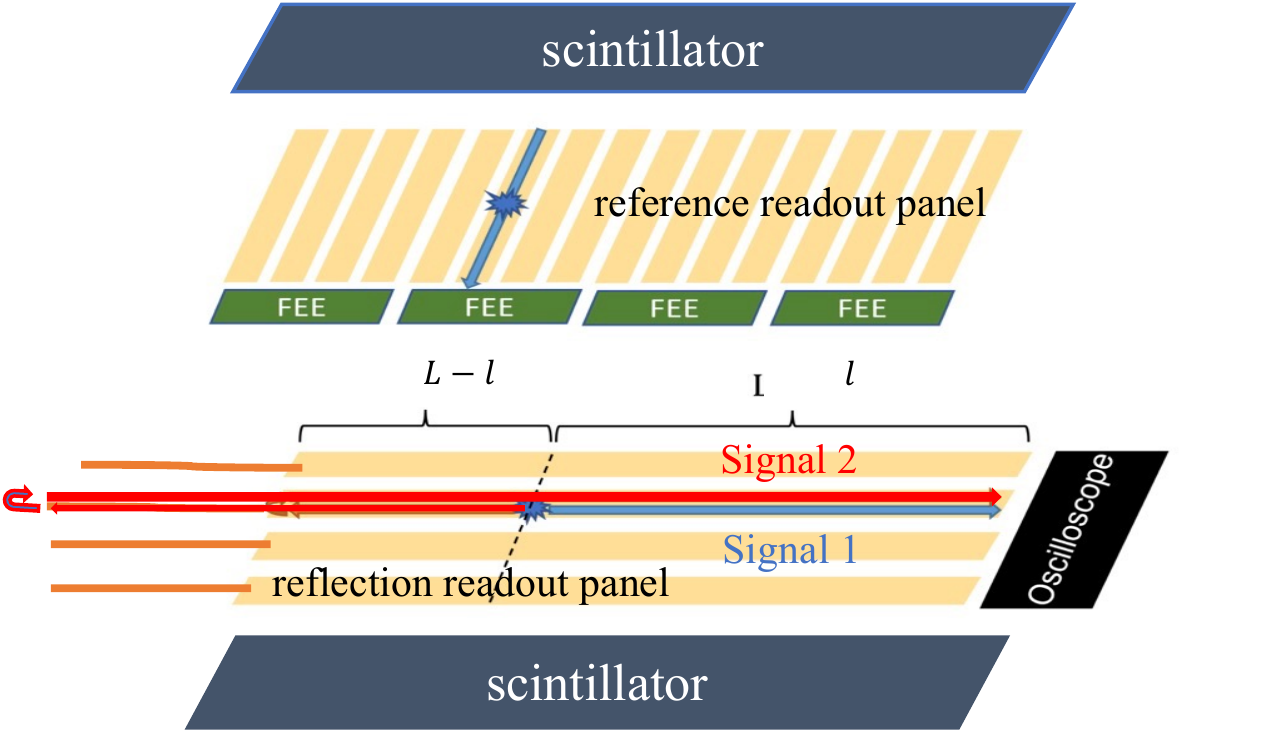} \label{X}}
	\qquad
	\subfloat[]{\includegraphics[width=2.3 in]{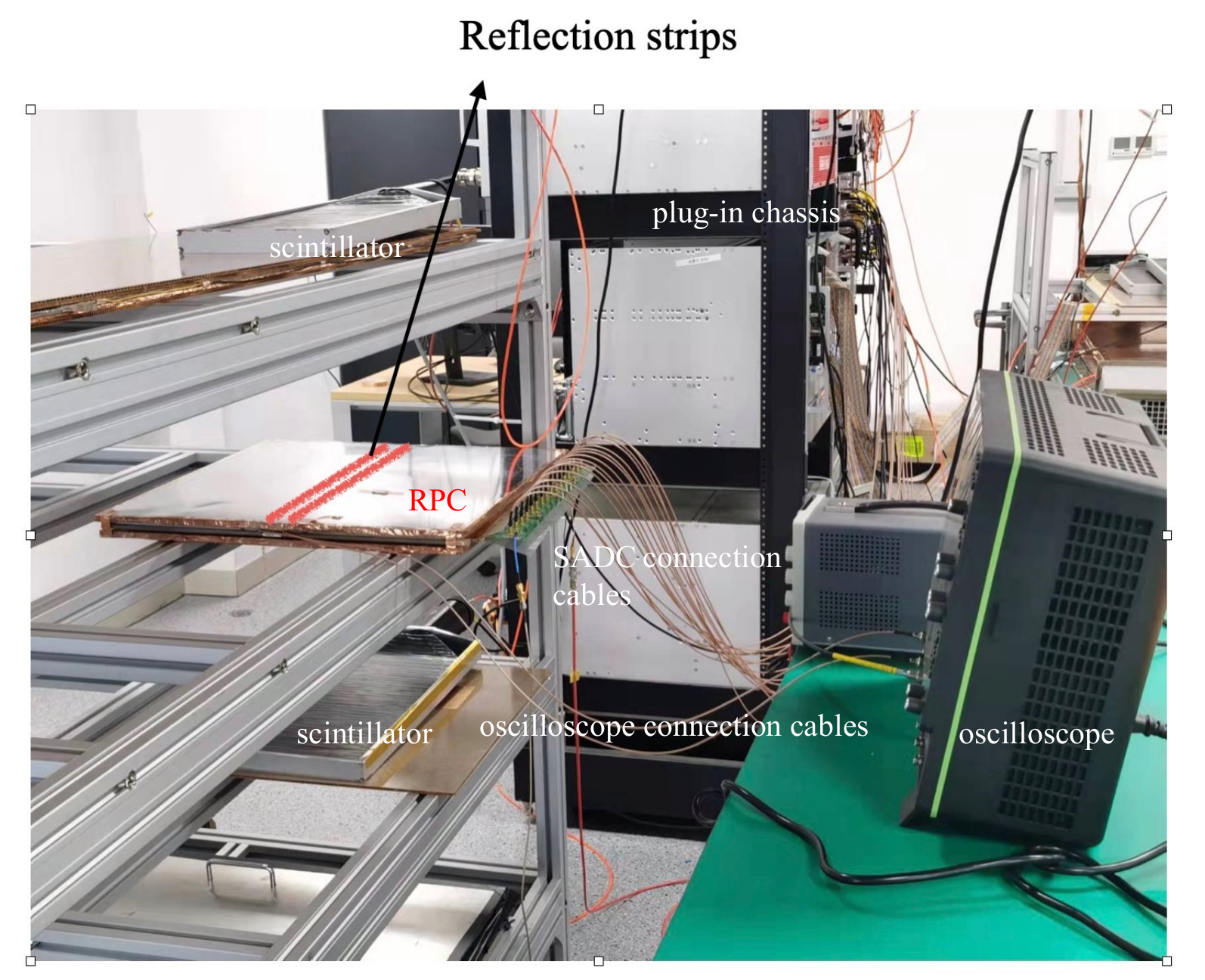} \label{Y}}
	\caption{(a) Side view of the experimental setup. (2) Photo of the experimental setup in laboratory.}
	\label{fig:setup}
\end{figure*}

To verify the feasibility of the scheme, one RPC prototype of 50 cm x 50 cm, made of 1-mm-thick bakelite gas gap and two orthogonal readout panels, is tested in a cosmic ray setup in our laboratory, as illustrated in Figure \ref{fig:setup}. This RPC is operated in avalanche mode with a gas mixture of tetrafluorethane (C$_2$H$_2$F$_4$), isobutane (Iso - C$_4$H$_{10}$) , and sulphur hexafluoride (SF$_6$) in proportions of 94.7$\%, 5.0\%$, and $0.3\%$, respectively.

Two readout panels are made up of the copper shield, 3-mm-thick foam ﬁlling and readout strips. There are 16 strips with a pitch of 27 mm on each panel, and the length of each strip is 50 cm. The upper panel is used as the reference readout panel which provides the reference hit position. The reference signals are read out via the amplifier boards, recorded by CAEN digitizers (V1742) and stored as waveforms with 1024 samplings for each pulse~\cite{Li_2021}. The sampling frequency is 5 GS/s. The bottom panel is used to test the reflection readout method. Six of the strips are equipped with the transmission cables. Length of each cable is 40 cm. The characteristic impedance is 20  $\Omega$, which is the measured impedance of such RPC structure~\cite{1697385}. In order to maintain integrity, Signal 1 and Signal 2 are digitised and recorded by the oscilloscope without ampliﬁers, and the sampling rate and bandwidth of the oscilloscope are respectively 20 GS/s and 1 GHz. 

Four scintillators of 40 cm x 20 cm are used for trigger. Two of them are positioned above the RPC while the other two are positioned below it. On each layer, the two scintillators are overlapped to make a coverage of 36 cm x 14 cm over the twelve middle strips of the reference panel and the six strips on the reflection panel.

\begin{figure*}[!h]
	\centering
	\subfloat[]{\includegraphics[width=2.4 in]{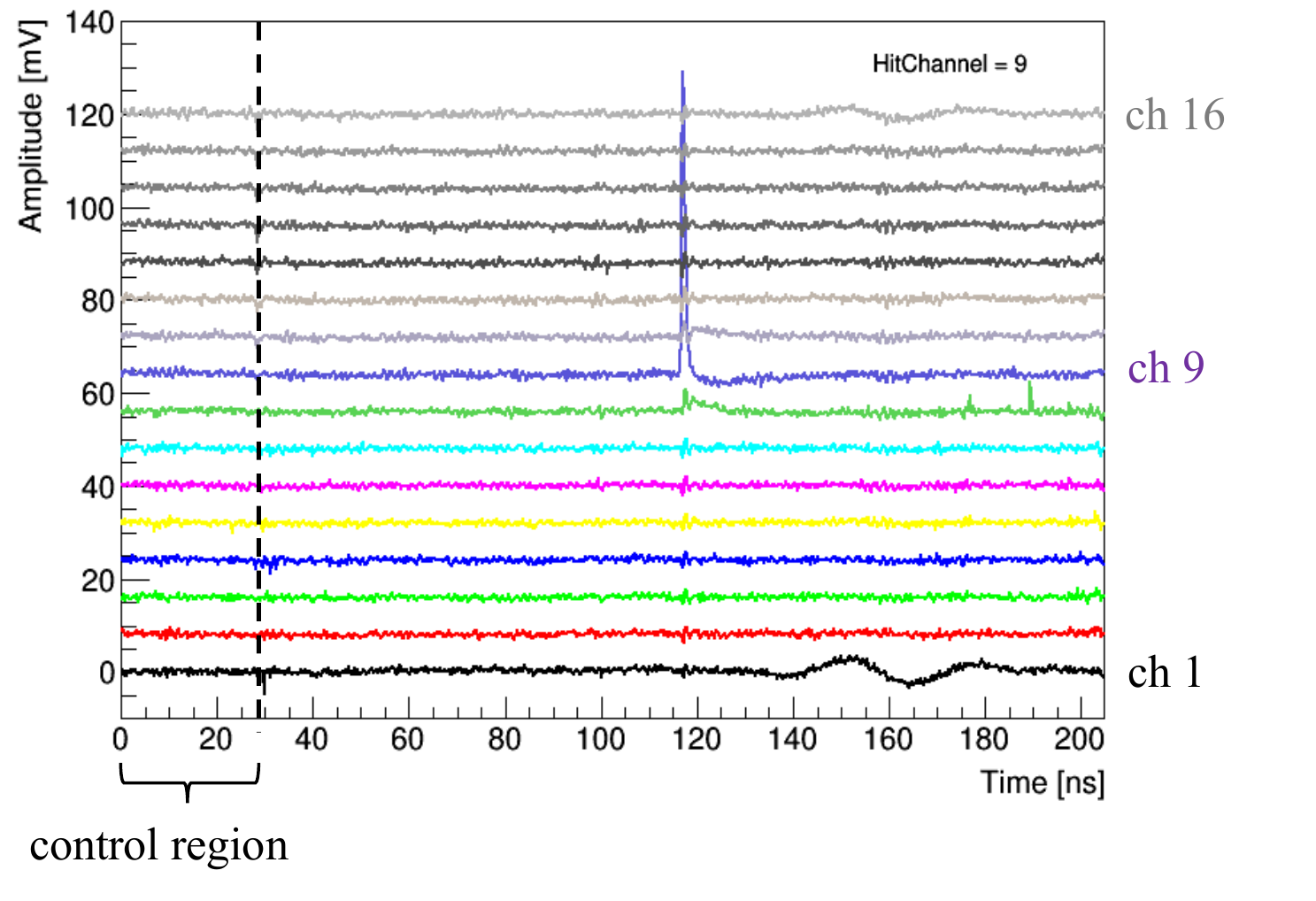} \label{X3}}
	\qquad
	\subfloat[]{\includegraphics[width=2.5 in]{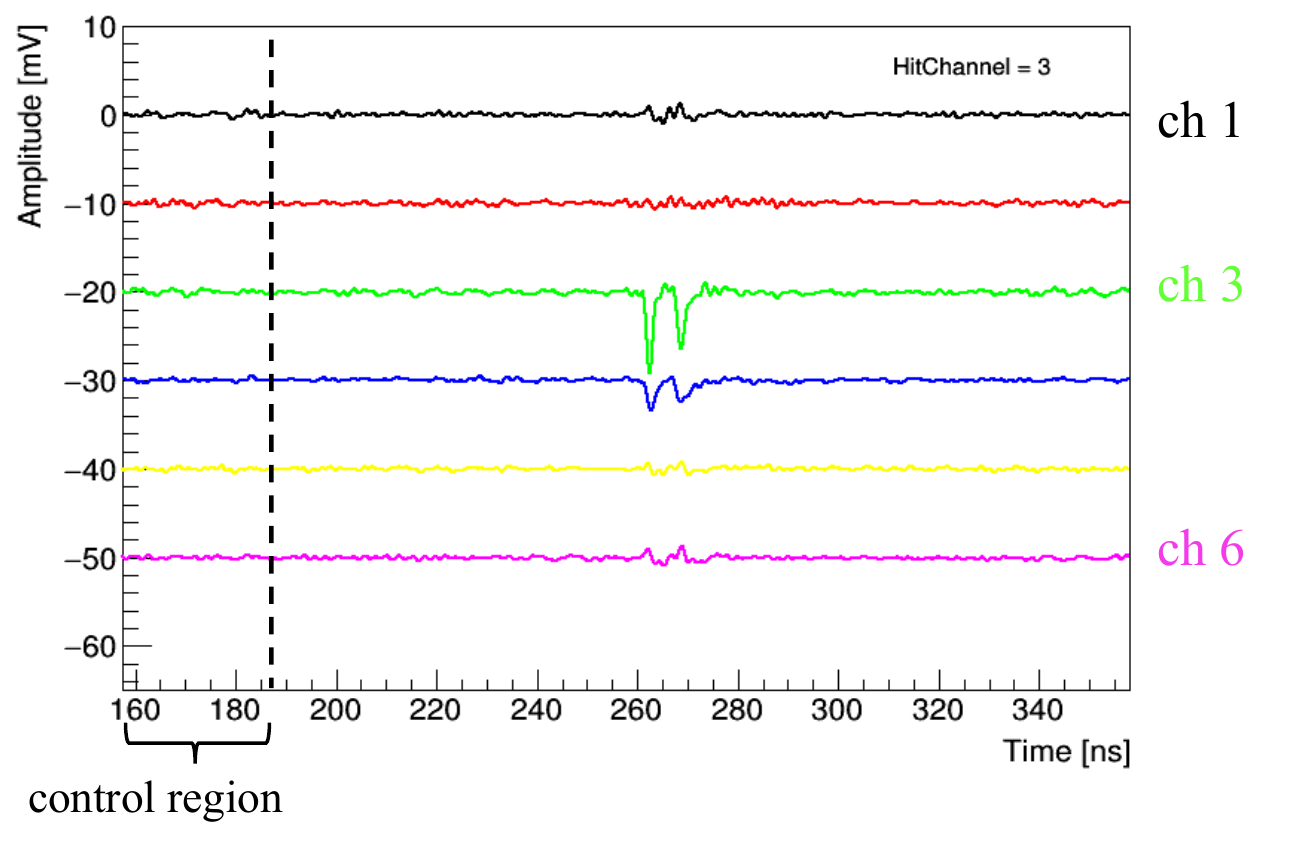} \label{Y3}}
	\caption{The waveform of a typical event (a) on the reference reaout panel. (b) on the reflection readout panel.}
	\label{fig:label4}
\end{figure*}

\section{Event selection}

A typical event recorded by the oscilloscope and digitizers is shown in Figure \ref{fig:label4}. Figure \ref{fig:label4}\subref{X3} shows the waveform of the positive signal on the reference readout panel. The time window from 0 ns to 30 ns, which is clearly out of the signal region (30 - 160 ns), is defined as the control region for the evaluation of the noise level of the measurement. Five times root mean square error of the waveforms in the control region is used as the signal discrimination threshold. The channel with the largest peak value in the signal region is defined as hit channel, and the center of this strip is used as the reference position. The events with more than one muon hitting two non-adjacent strips are excluded.

A waveform with two negative signals on channel 3 of the reflection readout panel is shown in Figure \ref{fig:label4}\subref{Y3}. Signal discrimination threshold of Signal 1 and Signal 2 is still the 5 times root mean square of the waveforms in the control region (from 0 ns to 30 ns). Reading from the figure, the two signals can be clearly distinguished and the time difference is about 15 ns.

Time difference of the two signals varies with the timing methods. In the following analysis, 70$\%$ peak value of the two signals was chosen as the nominal threshold to calculate the time difference. Other thresholds will be considered in Section 5.3 for position reconstruction and the spatial resolution calculation.

\section{Test results }
In order to verify the feasibility of the optimized reflection readout method, four aspects need to be considered: whether 'dead area' has been eliminated, efficiency of reflection readout, spatial resolution, signal attenuation.

\subsection{'dead area'  test}

Top priority of the feasibility study is to verify that the transmission cables should have eliminated the 'dead area'. In this test, the scintillators are moved on top of the far end of the readout strips. In this region, the time difference between Signal 1 and 2 is minimum. After connecting the 40 cm transmission cables, the time differences are always greater than 4 ns, which is much larger than the FWHM of a single RPC signal ($\sim$1.5ns). The original and reflected signal can be well separated. This shows that there is no 'dead area'. 

\subsection{Efficiency}

Efficiency is defined by fraction of events with both direct and reflected pulses recorded in all triggered events. Meanwhile, in order to check the efficiency loss from the readout method itself, another concept is introduced — reconstruction rate — which is defined as the fraction of events with both original and reflected signals in all triggered events that have also signals recorded on the reference panel.

\begin{figure}[!t]
	\centering
	\includegraphics[width=2.5 in]{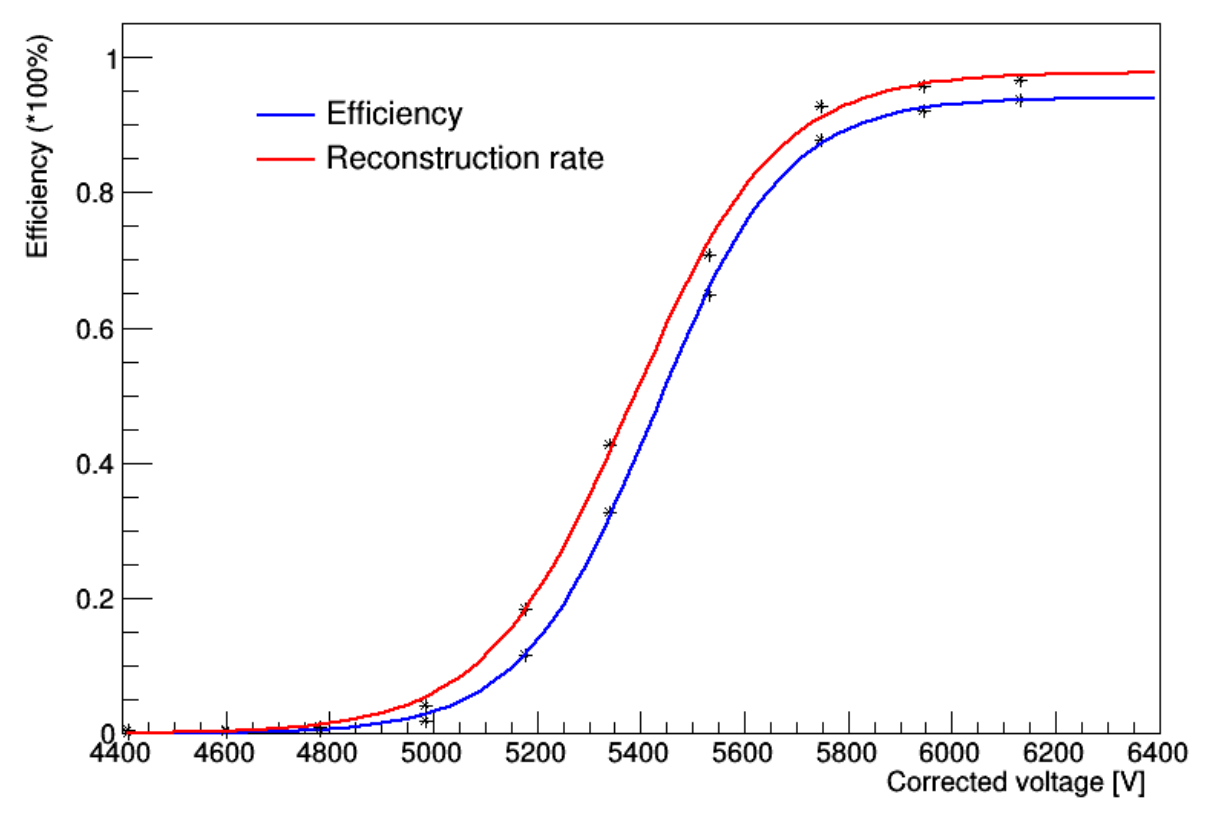}
	\caption{Efficiency and reconstruction rate distribution under different high voiltages.}
	\label{fig:label5}
\end{figure}

The efficiency and reconstruction rate is shown in Figure \ref{fig:label5}. The high voltage has been corrected to the normal pressure and temperature~\cite{Polini:2014usa}. The overall efficiency is about 92$\%$ taking no account of the reference signal. The reconstruction rate can be up to 97\%. In the following tests, the applied high voltage is set to 6000 V.

\subsection{Propagation velocity and spatial resolution}

Spatial resolution is the focus of this study, which can be evaluated from the difference between the reconstructed position with reflection readout method and the position measured by the orthogonal reference strips. The reference position is defined by the center of the strip. The reconstruction of the hit position on reflection readout panel need not only the time difference of two signals but also the propagation velocity of the signals along the readout strips, which can be achieved with the help of the reference position.

\begin{figure}[!b]
	\centering
	\includegraphics[width=2.5 in]{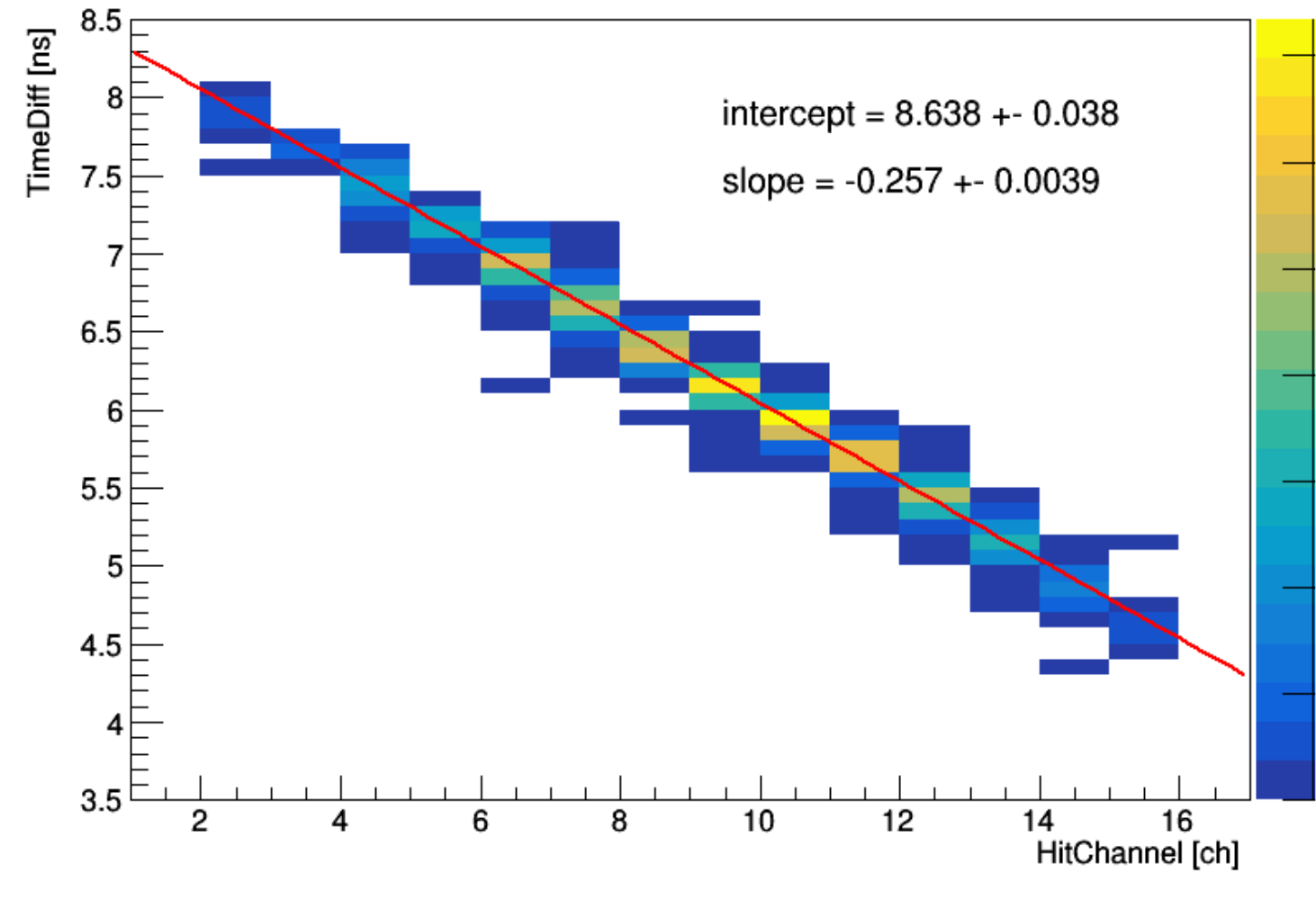}
	\caption{The relationship diagram between the time difference of two signals and the reference hit channel. Propagtion velocity of RPC signals on the readout panel can be calculated by the slope of the fitted line.}
	\label{fig:label6}
\end{figure}

Figure \ref{fig:label6} shows the linear relationship between the time difference of two signals and the reference hit channel. The arrival time is defined by the 70\% of the amplitude constant fraction discrimination (CFD). The propagation velocity can be given by $v=(2*l_{ch})\, / \, s,$ where $l_{ch}$ is the strip pitch of 27 mm, $s$ is the slope of the fit line. The calculated value is 210.1±3.2 mm/ns.

\begin{figure}[!t]
	\centering
	\includegraphics[width=2.6 in]{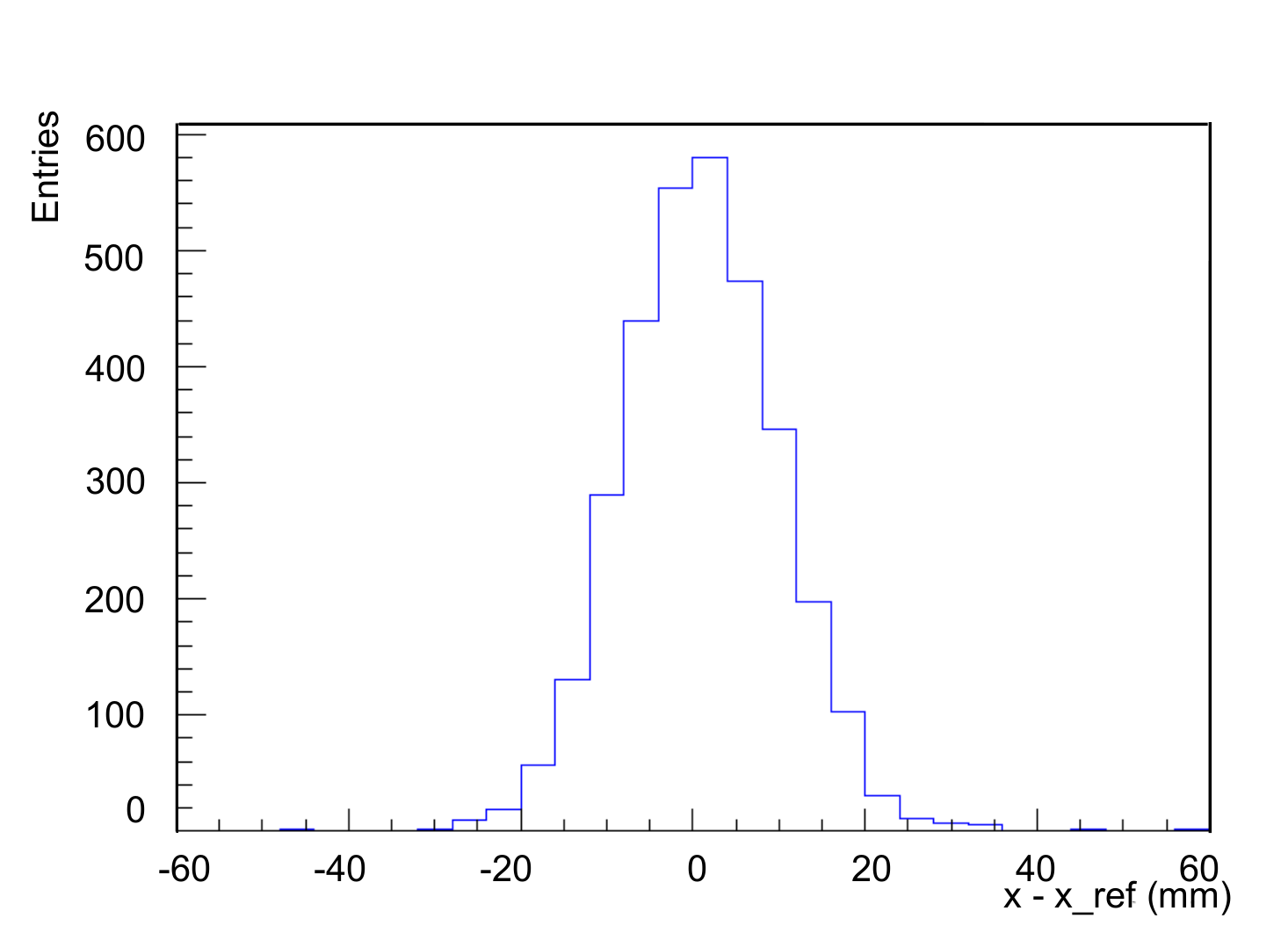}
	\caption{Distribution of the difference between the position determined with reflection readout method and that measured by the reference strips.}
	\label{fig:label7}
\end{figure}

The offset of the reconstructed position with respect to the reference position is presented as a histogram in Figure \ref{fig:label7}. The spatial resolution of the reflection readout method $\sigma_P$ is estimate with $$\sigma_P = \sqrt{\sigma_{Diff}^2 - (l_{ch}/\sqrt{12})^ 2},$$ where $\sigma_{Diff}$ is the root mean square of the offset distribution.

The position resolution is defined by the timing resolution. Since the waveforms of the signals are collected with the oscilloscope on the reflection readout panel, different timing methods can be applied.  The time of arrival is defined by CFD. Five different fractions, 50\%, 60\%, 70\%, 80\% and 90\% are considered. To find the time when the waveform crosses the CFD threshold, interpolation with a straight line is used to connect adjacent data points. The spatial resolutions of the reflection readout method under these different thresholds are separately 6.4 ± 0.1 mm, 5.9 ± 0.1 mm, 5.5 ± 0.1 mm, 5.2 ± 0.1 mm and 4.9 ± 0.1 mm. The best spatial resolution is achieved with CFD of 90\%. The resolution is about 5 mm. 

Besides the CFD method, two other common methods are also applied -- leading edge discrimination and zero crossing discrimination~\cite{NELSON2003324}. For the leading edge discrimination method,  -2 mV is chosen as the fixed threshold. The spatial resolution of this method is  8.4 ± 0.2 mm. For the zero crossing discrimination method, the time at the peak value of the signal is used and the spatial  resolution result is 6.1 ± 0.1 mm. With these less-performing timing methods, a position resolution better than 1 cm can still be achieved. 

\subsection{Signal attenuation}
When transmitted on the readout panels and transmission cables, the signal is attenuated on account of the wire resistance and dielectric loss. The surface resistivity of the graphite layers of the RPC gas gap is 1.2$\times$$10^5$ $ \Omega$/$\square$. The attenuation rate for the readout strip has been measured to be 14\%/m \cite{Xie_2021} previously.

To measure the attenuation of RPC signals transmitting along the transmission cables, three transmission cables of different lengths of 5, 40 and 80 cm are applied. The ratio of the amplitudes of the reflected and direct signals are measured to infer the attenuation rate. The ratio is plotted as a function of the cable lengths in Figure 7, from which the attenuation rate is found to be 8\%/m. 

As a cross check, the attenuation rate of the transmission cable is also measured with a network analyzer for sinusoidal signals with frequencies varied from 75 Hz to 325 Hz. The measured attenuation rate is in the range of 3.9\%/m to 9.7\%/m, which is consistent with the results above. 

\begin{figure}[!h]
	\centering
	\includegraphics[width=2.4 in]{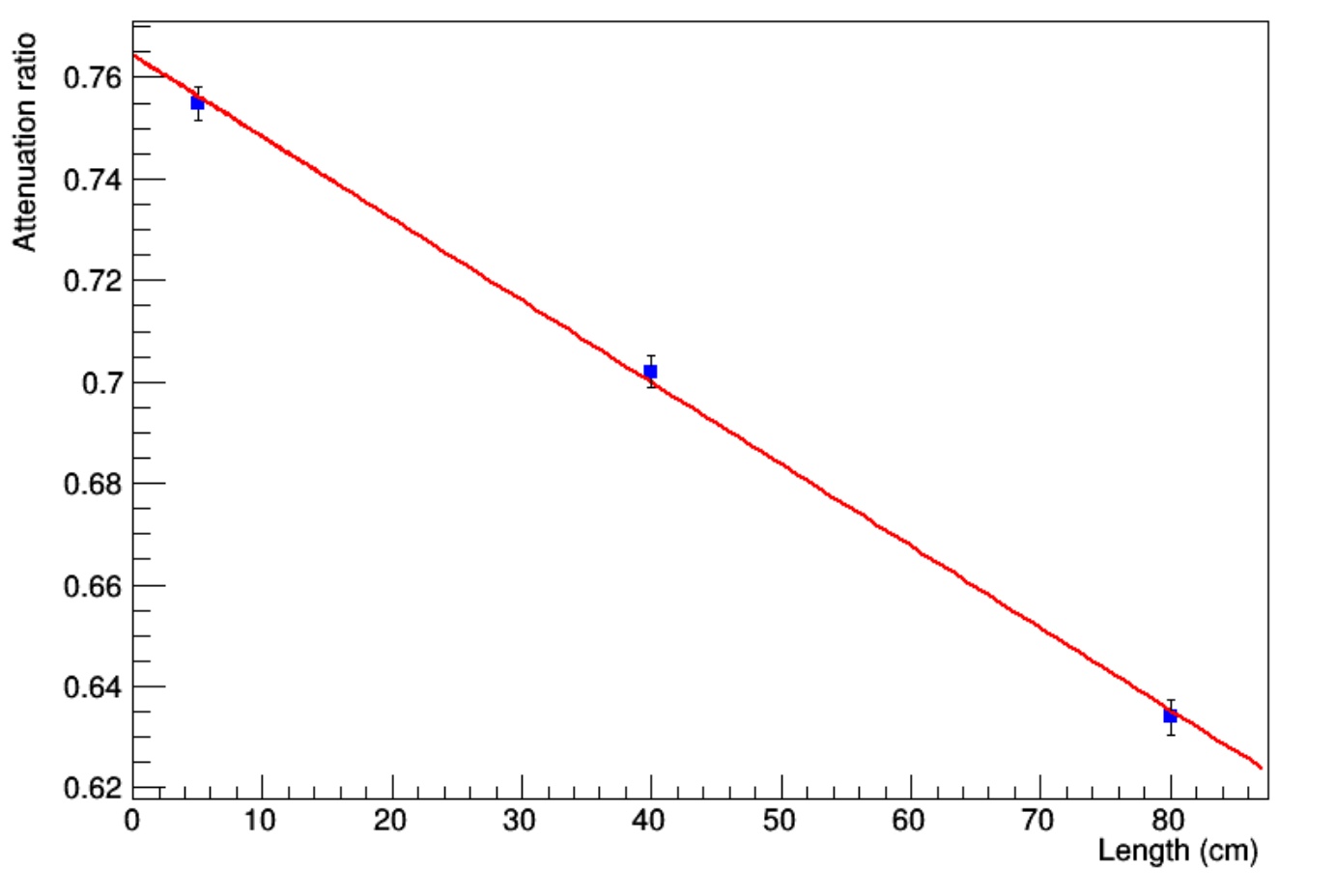}
	\caption{The ratio of the amplitude of the reflection and direct signals as a function of the length of the transmission cable.}
	\label{fig:label8}
\end{figure}

\section{Conclusion}
In this paper, reflection readout scheme is proposed and experimentally verified. With the transmission cable, the 'dead area' is suppressed. Reconstruction rate of the reflection readout method can be up to 97$\%$. The spatial resolution is less than 5 mm for delicate CFD method and better than 1 cm with fixed-threshold discriminator. These results strongly indicate the feasibility of the novel readout method, with highly reduced electronic channels, good spatial resolutjion and further improved geometrical acceptance.



\section*{Acknowledgements}
This work is partially supported by the National Natural Science Foundation of China (No. 11961141014).
\nocite{*}
\bibliographystyle{elsarticle-num}
\bibliography{double-end}



	
	
	

\end{document}